\documentclass[nofootinbib,aps,pra,superscriptaddress,onecolumn, titlepage=false]{revtex4-2}

\usepackage{graphicx}
\usepackage{amsmath}
\usepackage{mathrsfs}
\usepackage{dsfont}
\usepackage{amssymb}
\usepackage{subfigure}
\usepackage{color}
\usepackage{blkarray}
\usepackage{hyperref}
\usepackage{verbatim}
\usepackage{amsthm}
\usepackage{enumerate}
\usepackage{bbm}
\usepackage{hyperref}
\usepackage{braket}
\usepackage{booktabs}
\usepackage{mathtools}
\usepackage{titlesec}
\usepackage{array,booktabs}

\newcommand{\virg}[1]{``#1''}
\newcommand{\eq}[1]{Eq.~\eqref{#1}}
\newcommand{\cit}[1]{~\cite{#1}}

\newcommand*{\h}{\mathcal{H}}
\newcommand*{\Tr}{\mathrm{Tr}}

\newcommand*{\ti}{~}

\newcommand*{\coloneqq}{\mathrel{\vcenter{\baselineskip0.5ex \lineskiplimit0pt \hbox{\scriptsize.}\hbox{\scriptsize.}}} =}

\begin{document}
		\title{\Large Bulk area law for boundary entanglement in spin network states: entropy corrections and horizon-like regions from volume correlations}
		
	\author{Goffredo Chirco}
	\email{goffredo.chirco@unina.it}
	\affiliation{Dipartimento di Fisica ``Ettore Pancini'',
Universit\`a di Napoli Federico II, Napoli, Italy; INFN, Sezione di Napoli}
		
	\author{Eugenia Colafranceschi}
	\email{eugenia.colafranceschi@nottingham.ac.uk}
	\affiliation{School of Mathematical Sciences and Centre for the Mathematics and Theoretical Physics of Quantum Non-Equilibrium Systems, University of Nottingham, University Park Campus, Nottingham NG7 2RD, United Kingdom}
	
	\author{Daniele Oriti}
	\email{daniele.oriti@physik.lmu.de}
	\affiliation{Arnold Sommerfeld Center for Theoretical Physics, \\ Ludwig-Maximilians-Universit\"at München \\ Theresienstrasse 37, 80333 M\"unchen, Germany}

\begin{abstract}
For quantum gravity states associated to open spin network graphs, we study how the entanglement entropy of the boundary degrees of freedom (spins on open edges) is affected by the bulk data, specifically by its combinatorial structure and by the quantum correlations among intertwiner degrees of freedom. For a specific assignment of bulk edge spins and slightly entangled intertwiners, we recover the Ryu-Takayanagi formula (with a properly (discrete) geometric notion of area, thanks to the underlying quantum gravity formalism) and its corrections due to the entanglement entropy of the bulk state. We also show that the presence of a region with highly entangled intertwiners deforms the minimal-area surface, which is then prevented from entering that region when the entanglement entropy of the latter exceeds a certain bound. This entanglement-based mechanism leads thus to the formation of a black hole-like region in the bulk.
\end{abstract}
	\maketitle

	\section*{Introduction}		
	
	 Entanglement is expected to play a crucial role in the emergence of spacetime and geometry from fundamental quantum entities, and to account for such emergence is a key goal of all quantum gravity approaches. In fact, a connection between entanglement and geometry, both at the kinematic and dynamical levels, has been pointed out in several contexts~\cite{Ryu:2006bv, Ryu:2006ef, VanRaamsdonk:2009ar,VanRaamsdonk:2010pw,lashkari2013gravitational,Faulkner_2014,Cao:2016mst}, in a semi-classical as well as in full quantum gravity regimes. An important example of this connection is provided by the Ryu-Takayanagi formula~\cite{Ryu:2006bv,Ryu:2006ef}, which relates, in the AdS/CFT context~\cite{Hubeny:2014bla}, the entanglement entropy of a boundary region to the area of a minimal surface in the dual bulk. In fact, many of the recent results connecting entanglement measures with geometric quantities have been obtained taking advantage of the holographic correspondence, since this is among the best examples we have of reconstruction of (some aspects of) geometry and gravitational physics from non-gravitational one (although based on a background spacetime, i.e. the flat boundary of AdS). However, it is important to point out that, just like holography is expected to be a fundamental principle of quantum gravity way beyond the specific AdS/CFT realization, the entanglement/geometry correspondence is also expected to be realized in quantum gravity independently of holography (and AdS/CFT). The idea of entanglement as \virg{the fabric of spacetime}~\cite{Swingle:2009bg} suggests to think at a theory of quantum gravity in terms of fundamental entities glued together by entanglement. 
	 
	 This picture is common to all quantum gravity formalisms in which spacetime and geometry are emergent, and in such context the entanglement/geometry correspondence is basically a logical necessity, not a conjecture. Tensorial group field theory~\cite{Oriti:2011jm,Krajewski:2012aw,Oriti:2014uga,Rivasseau:2012yp,Carrozza:2016vsq, Rivasseau:2016zco, Rivasseau:2016wvy, Delporte:2018iyf} (TGFT), and even more clearly the class of TGFT models with richer quantum geometric ingredients, called simply {\it group field theory}~\cite{Oriti:2011jm,Krajewski:2012aw, Oriti:2013aqa, Oriti:2014uga} (GFT) provides such a formalism. It  describes spacetime as a collection of fundamental quantum simplices (to be understood as quanta of space) that, upon becoming entangled, constitute discrete spatial geometries, and by interacting give rise to spacetime manifolds of arbitrary topology. The same fundamental entities and the same overall picture is shared by closely related quantum gravity approaches like canonical loop quantum gravity~\cite{Ashtekar:2021kfp,Thiemann:2007zz} and spin foam models~\cite{Perez:2012wv,Freidel:2007py}. 
	 
	 In this picture, quantum spacetime is thus described as a many-body system, whose structure of entanglement (and of quantum correlations more generally) determines its geometry and topology. Quantum information tools are thus necessary to study such a system and to extract geometric information from its quantum data; in particular, as we are dealing with entanglement in quantum many-body systems, tensor networks~\cite{Ran_2020,2006quant.ph..8197P, ORUS2014117,Bridgeman_2017} (TN) are the ideal means to approach the problem, as it is the case also in the cited AdS/CFT context. In this work, we study entanglement properties of spin network states, shared by all the mentioned quantum gravity formalisms, and make extensive use of tensor network techniques to study some of their entanglement properties.
	 
	 In~\cite{Colafranceschi:2020ern}, building also on previous work~\cite{Chirco2018a, 2020}, a precise correspondence between classes of quantum gravity states and symmetric tensor networks has been established. A direct application of that  dictionary is then provided by \cite{Colafranceschi:2021acz} where, for regions of quantum space, the properties of transmission of information from the bulk to the boundary, for a class of states corresponding to random tensor networks, is explored. 
	 
	 Here we investigate further the relation between bulk and boundary of quantum gravity states corresponding to spatial regions in terms of entanglement entropy. We analyse how the entanglement entropy of the boundary is affected by the quantum correlations of the bulk (and its entanglement entropy). In particular, we investigate under which conditions the Ryu–Takayanagi entropy formula holds in this context, and how the bulk entanglement modifies it, computing the corresponding corrections. Next, we explore the features of the quantum states that affect such entropy formula, and identify the cases in which it reproduces a black hole-like (better, horizon-like) situation, characterized in purely quantum information terms (in other words, we do not explore all the geometric aspects).
	 
Our work is a generalization of what has been done in the context of random tensor networks by Hayden \textit{et al.} in\cit{Hayden:2016cfa}, with a crucial (with respect to the quantum gravity interpretation) change in perspective: the tensor networks we work with inherently possess a quantum-geometry characterization, being dual by construction to a triangulation of quantum space (as opposed to those used within the tensor networks/AdS correspondence~\cite{Swingle:2009bg}, where such a characterization is implemented at a later stage, thanks to a definition of the metric in combinatorial terms). Crucially, this implies that the corrections to the Ryu–Takayanagi entropy formula we find derive from the quantum-geometric properties of the bulk, and thus differ from that obtained in the AdS/CFT context (see, for example,~\cite{swingle2014universality}), which are related to the semiclassical description of the bulk geometry. The present results also differ from previous work in~\cite{2020}, where corrections to the area-law were directly associated to  perturbations of the free GFT model. 


Let us finally mention that holographic bulk/boundary mapping in spin network states has recently been investigated also in\cit{Chen:2021vrc}. Such interesting work, however, focuses on a bulk reconstruction from the boundary density matrix, thereby following a different approach from the one used here, where properties of the bulk/boundary mapping are investigated through entanglement entropy evaluation.

\section{Framework}

In this section, we introduce the main features of the quantum gravity states we are going to analyze in the following. We only deal with kinematical properties and therefore do not discuss the quantum dynamics these states are subject to. We adopt the GFT formulation of the Hilbert space of such states, and a second quantized language based on a Fock space, since the way combinatorial structures associated to quantum gravity states encode entanglement correlations of the fundamental degrees of freedom is most manifest in this formalism. In doing so, we focus on the single-body Hilbert space first, explaining then how such Hilbert space, associated to a single spin network vertex (dual to a 3-simplex), is used to define states associated to extended structures, i.e. generic spin network states associated to arbitrary complex graphs (dual to gluings of 3-simplices) labelled by group representations. It is important to stress, however, that the very same spin network states are shared also with canonical loop quantum gravity and spin foam models, and that, in fact, as long as one considers only quantum states associated to a given graph, there is no difference between these formalisms and they all use the same Hilbert space of quantum states\cit{Oriti:2014uga}. Since this is the case in our present analysis, our results are going to be equally valid in all these quantum gravity formalisms, and can be seen (and further developed) from the perspective of each of them. 

We focus on quantum spatial geometries, which are described in GFT as collections of spin network vertices with  maximal entanglement of edge-spins expressing links between the respective vertices (adjacency relations between the dual simplices). For such states, then, the degrees of freedom of edges forming links are partially \virg{frozen}, as they have been projected into maximally entangled states for a given irreducible representation associated to the edge, which remains dynamical. The relevant group is in principle arbitrary, but quantum gravity models make use mostly of $SU(2)$ data, which will be also our choice in the following. The remaining degrees of freedom are group intertwiners attached to the vertices (we consider them \textit{bulk} degrees of freedom) and spins (more generally, group irreps) attached to the open edges (we consider them \textit{boundary} degrees of freedom). In the next subsection we present all this in more mathematical detail. We are then going to explore how the entanglement entropy of the boundary is affected by the bulk, specifically by its combinatorial structure (generated by entanglement on edge spins) and by its quantum amplitude, i.e. the quantum correlations among intertwiners. 

The class of states we consider can be expressed equivalently as random tensor networks. Every vertex has a random wave-function attached to it, and the spin network graph is realized by projecting appropriate pairs of edge spins onto maximally entangled states. That is, we work with random symmetric PEPS tensor networks. Also this correspondence will be explained in some more detail in the next subsection. We then insert quantum correlations between the intertwiners of the resulting network/graph and study how they affect the entanglement entropy of the boundary\footnote{In other words, we project the random tensor network on a certain bulk state and study the entropy of the resulting boundary state.}.

\subsection{Spin networks from entanglement of GFT vertices}

\begin{figure}
	\centering
	\includegraphics[width=0.25\linewidth]{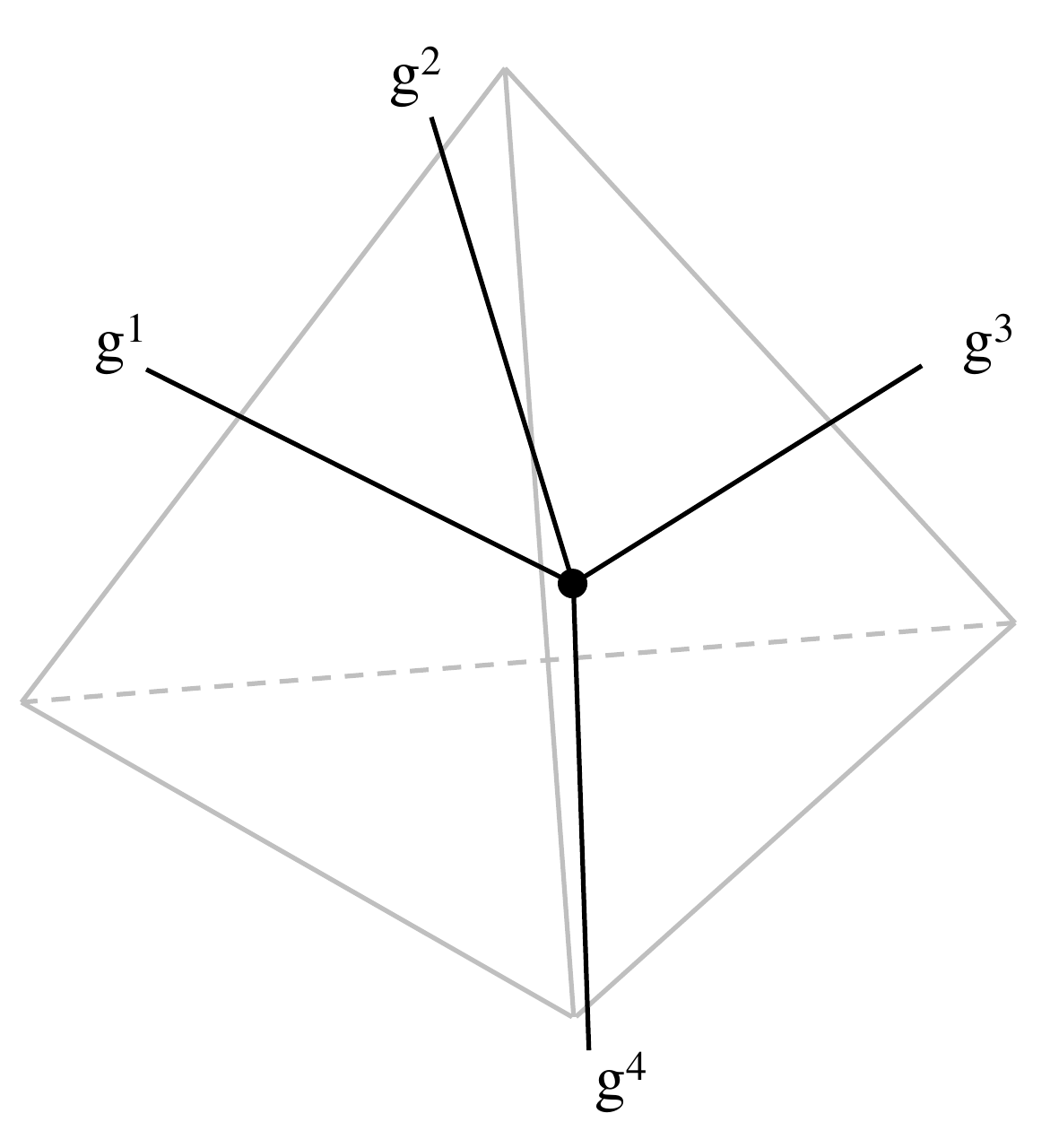}
	\caption{A four-valent vertex (depicted in black) dual to a tetrahedron (depicted in a brighter colour). The set of edges, each one dual to a face of the tetrahedron, is decorated with an equivalence class of group variables: $[g^1,g^2,g^3,g^4]=\{hg^1,hg^2,hg^3,hg^4|h\in G\}$.}
	\label{fig:vertex}
\end{figure}

In a single GFT vertex (i.e. a spin network vertex) each edge carries a colour $i=1,...,d$ and is decorated by a group variable $g^i\in G$. Here and in the following, we assume $G=SU(2)$. In order to correctly describe the discrete geometry of a simplex, this structure has to be gauge invariant, i.e. invariant under the simultaneous action of an arbitrary group element on all the edges. Its Hilbert space is thus given by $\h_v\coloneqq L^2(G^d/G)$. We show in figure \ref{fig:vertex} the example of a four-valent vertex which is dual to a tetrahedron.

By applying the Peter-Weyl theorem, a vertex wave-function $f \in L^2(G^d/G)$  with $G=SU(2)$ can be expanded in the spin network basis $\{\ket{\vec{j}\vec{n}\iota}\}$, where $\vec{j}=j^1...j^d$ are representations (spins) of $SU(2)$ and $\vec{n}=n^1...n^d$ indices labelling a basis in the corresponding representation spaces, e.g. $\ket{j^in^i}$ is a basis element of the $j^i$-representation space $V^{j^i}$; and $\iota$ is the intertwiner quantum number arising from the gauge-invariant recoupling of the edge spins: $\ket{\vec{j}\iota} \in \text{Inv}_{SU(2)}\left[V^{j^1}\otimes ... \otimes V^{j^d}\right]\coloneqq \mathcal{I}^{\vec{j}}$. In particular, we have that
\begin{equation}\begin{split}\label{f}
\ket{f}=\bigoplus_{\vec{j}}\sum_{\vec{n}\iota} f^{\vec{j}}_{\vec{n}\iota} ~ \ket{\vec{j}\vec{n}\iota} ~\in ~ \h_v=\bigoplus_{\vec{j}} \h_v(\vec{j}), \quad \text{with}\quad  \h_v(\vec{j})\coloneqq \mathcal{I}^{\vec{j}}\otimes \bigotimes_{i=1}^dV^{j^i}
\end{split}
\end{equation}

In this framework, graphs dual to simplicial complexes arise as collection of open vertices (dual to simplices) whose edges are glued together by entanglement, as we are going to show explicitly. Since the background independence of gravity requires the vertices to be indistinguishable in absence of additional dynamical variables (they cannot be localized in a manifold by coordinates, for example), the resulting spin network states live in the Fock space $\mathcal{F}(\h_v)=\bigoplus_N \mathrm{sym}\left( \h_v^{\otimes N}\right)$, where $N$ is the number of vertices. However, for practicality as well as for easier adapting of our procedure in the other related quantum gravity formalisms based on spin networks,  we adopt a first-quantized language, with the implicit assumption that all physically relevant quantities are symmetrized with respect to the vertex labels.

Let us start with a set of $N$ open vertices described by a state $\ket{\psi}$; the projection of the latter into a maximally entangled states of a pair of edge-spins creates a \textit{link} between the corresponding vertices. For example, edges $e_v^i$ and $e_w^i$ can be glued to form a link $e_{vw}^i$ thanks to the projection of $\ket{\psi}$ into the maximally entangled state
\begin{equation}
\ket{e_{vw}^i}\coloneqq \bigoplus_j\frac{1}{\sqrt{d_{j}}}\sum_n  \ket{j n }\otimes \ket{jn},
\end{equation}
where $d_j\coloneqq 2j+1$ is the dimension of the representation space $V^j$ to which $\ket{jn}$ pertains. More generally, spin network states with connectivity $\gamma$ can be obtained as follows:
\begin{equation}\begin{split}\label{gamma}
\ket{\psi_\gamma}=\left(\bigotimes_{e_{vw}^i \in L} \bra{e_{vw}^i}\right)\ket{\psi} 
\end{split}
\end{equation}
where $L$ is the set of internal links of $\gamma$.

\begin{figure}
	\centering
	\includegraphics[width=0.7\linewidth]{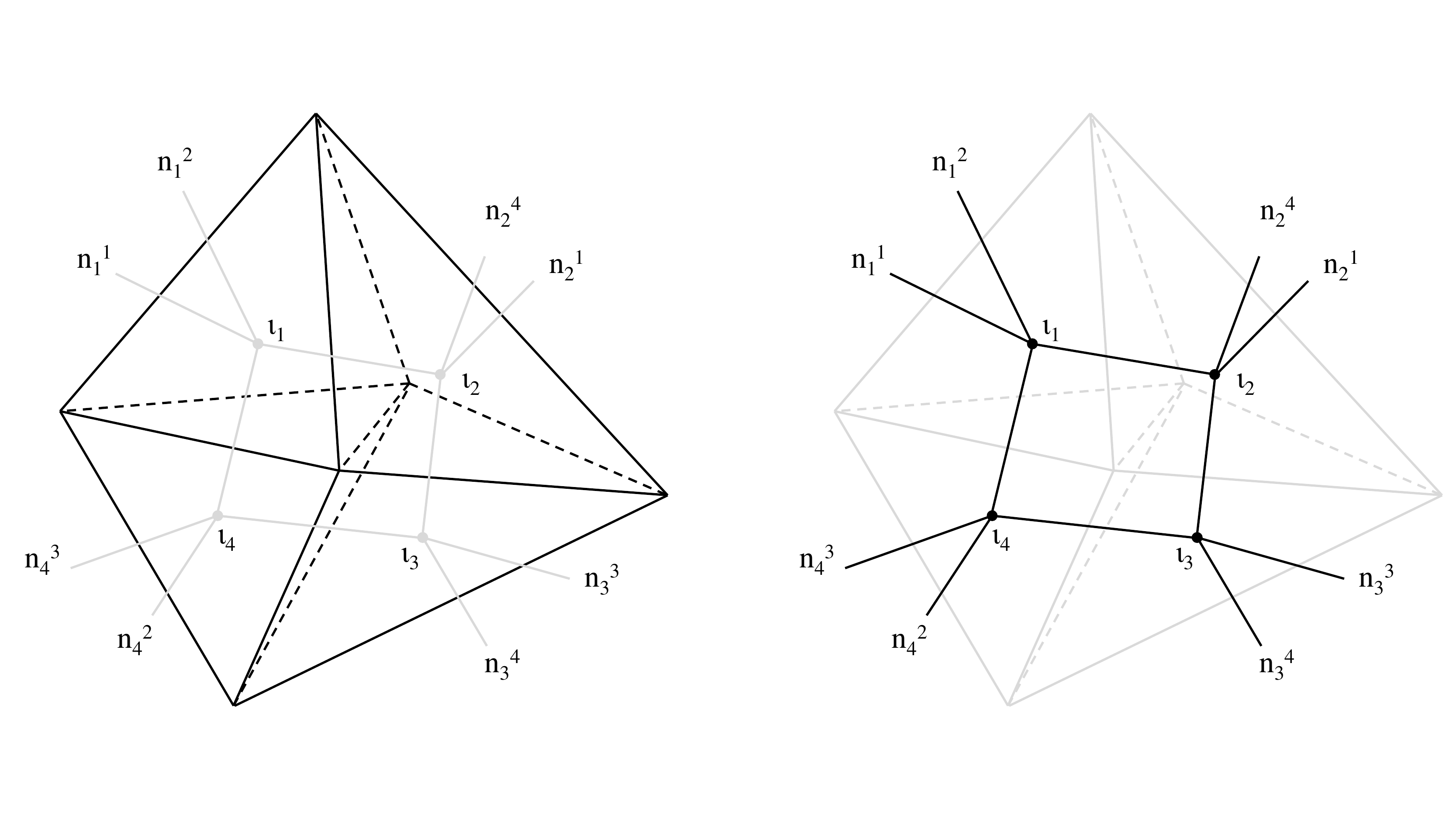}
	\caption{The gluing of simplices (each one dual to a spin network vertex) builds up a simplicial complex (marked in black on the left) dual to a spin network graph (marked in black on the right). For a given assignment of edge spins (which are omitted in the figure), the degrees of freedom of the resulting structure are intertwiners $\iota_v$ attached to the core of the vertices and spin projections $n_v^i$ attached to open edges.}
	\label{fig:presentazione2}
\end{figure}

We focus on a specific class of states, corresponding to random tensor networks. We start from a set of open vertices with individual weights $ (f_v)^{\vec{j}}_{\vec{n}\iota} $ peaked on a certain spin-set $ \vec{j} $, each one chosen randomly within the corresponding Hilbert space $ \h_v(\vec{j}) $, and glue them according to the connectivity pattern $\gamma$. The states we consider, therefore, possess the form
\begin{equation}\begin{split}
\ket{\tau_\gamma}=\left(\bigotimes_{e_{vw}^i \in L} \bra{e_{vw}^i}\right)\bigotimes_v \ket{f_v}
\end{split}
\end{equation}
and pertain to the Hilbert space
\begin{equation}
\h_{\gamma}(J) \coloneqq  \bigotimes_{v=1}^N \mathcal{I}^{\vec{j}_v}\otimes \bigotimes_{e_{v}^i\in \partial \gamma} V^{j_v^i}
\end{equation}
where $J$ is the set of spins on all links of $\gamma$. Note that, after the gluing, the set of vertices is left with the following degrees of freedom: spin labels and spin projections on open (non-contracted) edges, i.e. the \textit{boundary} of the spin network graph, and intertwiners labels attached to the vertices themselves, which constitute the \textit{bulk} (see figure \ref{fig:presentazione2}).  The graph Hilbert space $\h_{\gamma}(J)$ thus factorizes into a boundary Hilbert space $\h_{\partial \gamma}(J_{\partial })\coloneqq \bigotimes_{e_{v}^i\in \partial \gamma} V^{j_v^i}$ and a bulk Hilbert space $\h_{\dot{\gamma}}(J) \coloneqq \bigotimes_{v=1}^V \mathcal{I}^{\vec{j}_v}$ (note that the latter depends on all spins attached to the graph $\gamma$). \medskip

\subsection{Rényi entropy for random tensor network from Ising free energy}
We want to study how the entanglement entropy of the boundary of a spin network is affected by the bulk, specifically by its combinatorial structure and quantum correlations among intertwiners. To this end, we consider states of the form of \eq{gamma} and, upon regarding them as bulk-to-boundary maps  (following  \cite{Colafranceschi:2021acz}), we focus on the boundary state produced by a certain input $\ket{\zeta} \in\h_{\dot{\gamma}}(J)$ for the bulk; specifically, we look at the \virg{process}
\begin{equation}
\ket{\zeta} \quad \rightarrow \quad \ket{\tau_{\partial\gamma}(\zeta)}\coloneqq\bra{\zeta}\tau_\gamma \rangle,
\end{equation} 
where $\ket{\tau_{\partial\gamma}(\zeta)} \in  \h_{\partial \gamma}(J_\partial)$ is the boundary output state we focus on. In particular, we compute the Rényi-2 entropy of a portion $A$ of the boundary, which is given by 
\begin{equation}
S_2(\rho_A)=\Tr\left[\rho_A^2\right]
\end{equation} 
where $\rho_A$ is the reduced density matrix of $A$, i.e. $\rho_{A}\coloneqq\Tr_{\overline{A}}\left[\rho\right]$ with $\rho=\ket{\tau_{\partial\gamma}(\zeta)}\bra{\tau_{\partial\gamma}(\zeta)}$ and $\overline{A}$ being the set of boundary edges complementary to $A$. Using the replica trick, this can be computed as follows:
\begin{equation}
S_2(\rho_A)=-\log \left(\frac{Z_1}{Z_0}\right)\quad \text{with} \quad \begin{aligned}
&Z_1 \coloneqq \Tr \left[\left(\rho \otimes \rho\right)\mathcal{S}_A\right]\\&Z_0 \coloneqq \Tr \left[\rho \otimes \rho\right]
\end{aligned}
\end{equation}
where $\mathcal{S}_A$ is the operator swapping the two copies of portion $A$ of the boundary state.  Due to the random character of the vertex states we compute the average value of the entropy and consider the regime of large spins\footnote{For a generalization of the calculation to arbitrary bond dimensions, see \cite{Vasseur:2018gfy}.}, in which 
\begin{equation}\label{av}
\overline{S_2(\rho_A)}\simeq-\log \frac{\overline{Z_1}}{\overline{Z_0}},
\end{equation}
with
\begin{equation}\begin{split}
\overline{Z_{0}} &=\Tr \left[\rho_\zeta^{\otimes 2}~\rho_{L}^{\otimes 2}~\bigotimes_v \overline{\rho_{v}^{\otimes 2}}\right]\\ \overline{Z_{1}} &=\Tr \left[\rho_\zeta^{\otimes 2}~\rho_{L}^{\otimes 2}~\bigotimes_v \overline{\rho_{v}^{\otimes 2}}~\mathcal{S}_A\right]
\end{split}
\end{equation}
where $\rho_\zeta = \ket{\zeta}\bra{\zeta}$, $\rho_{L}\coloneqq \bigotimes_{e\in L}\ket{e}\bra{e}$ and $\rho_v = |f_v\rangle \langle f_v|$. As showed in \cite{Colafranceschi:2021acz} (by adapting to the quantum gravity framework the random tensor techniques of \cite{Hayden:2016cfa}), for a uniform probability distribution of the vertex wavefunctions the quantities $\overline{Z_{0}}$ and $\overline{Z_{1}}$ correspond to partition functions of a classical Ising model defined on the graph $\gamma$\footnote{We stress that the dual Ising model arises because we are computing the 2nd Renyi entropy and the values of the Ising spins refer to the two copies of the system entering this computation.}.  The key aspects of the calculation are the following:

\begin{itemize}
	\item[-] \textit{Randomization and Ising spins}
	
	The randomization yields an Ising spin $\sigma_v$ for every vertex $v$, which can be $+1$ or $-1$. Effectively, each edge $e_v^i$ of the vertex carries a copy of the spin $\sigma_v$. Note that all such copies have the same value; in fact, $\sigma_v$ expresses an \virg{Ising state} of the vertex, which is \virg{transmitted} to all its edges.  \smallskip 
	
	\item[-] \textit{Probed region and pinning fields}
	
	A set of virtual spins called \virg{pinning fields} keeps track of the boundary region $A \subset \gamma $ respect to which the entropy is computed. In particular, a spin $\mu_{e_v^i}$ is attached to every boundary edge $e_v^i \in \partial \gamma$, and takes value $-1$ if $e_v^i \in A$, and $+1$ otherwise. \smallskip
	
	\item[-] \textit{Interaction of Ising spins and pinning fields}
	
	The Ising spin $\sigma_v$ on a boundary edge $e_v^i$ interacts to the pinning field $\mu_{e_v^i}$ living on the same edge (with strength of the interaction proportional to $\log d_{j_v^i}$). Moreover, the Ising spin $\sigma_v$ on a semi-link $e_v^i$ interacts to the Ising spin $\sigma_w$ on the complementary semi-link $e_w^i$, and the strength of interaction is proportional to $\log d_{j_{vw}^i}$. Equivalently, we could say that the Ising spins (when regarded as pertaining to the vertex in its entirety, not split in copies attached to the vertex substructures) interact to their nearest neighbours: $\sigma_v$ interacts with $\sigma_w$ if $v$ and $w$ are connected by a link.\smallskip
	
\end{itemize}

\noindent
In particular, we can define the partition function 
\begin{equation}\begin{split}
&\overline{Z}\left(\vec{\mu}\right)\coloneqq\sum_{\vec{\sigma}}e^{-\mathcal{A}[\vec{\mu}]\left(\vec{\sigma}\right)},
\end{split}
\end{equation}
where $\mathcal{A}[\vec{\mu}](\vec{\sigma})$ is an Ising-like action, function of the Ising spins $\vec{\sigma}=\{\sigma_v|v=1,...,N\}$ and with a parametric dependence (expressed by the square brackets) on the boundary pinning fields $\vec{\mu}=\{\mu_v^i|e_v^i \in \partial \gamma\}$:
\begin{equation}\begin{split}
\mathcal{A}[\vec{\mu}](\vec{\sigma})&= -\frac{1}{2}\sum_{e_{vw}^i\in L}(\sigma_v\sigma_w -1)\log d_{j_{vw}^i}
-\frac{1}{2}\sum_{e_{v}^i\in \partial \gamma}(\sigma_v\mu_{e_v^i}-1)\log d_{j_v^i} + S_2({\rho_\zeta}_\downarrow)+ \mathrm{const}
\end{split}
\end{equation}
with $S_2({\rho_\zeta}_\downarrow)$ Rényi-2 entropy of the bulk state reduced to the region with Ising spins $\sigma_v=-1$. We then have that
\begin{equation}\begin{split}
&\overline{Z_0}=\overline{Z}\left(\mu_{e}=+1~ \forall ~ e \in \partial \gamma\right), \qquad \overline{Z_1}=\overline{Z}\left(\mu_{e}=-1~ \text{if} ~ e \in A, \mu_{e}=+1~ \text{if} ~ e \notin A\right)
\end{split}
\end{equation}
From the partition function $\overline{Z}\left(\vec{\mu}\right)$ we can define the free energy $F\left(\vec{\mu}\right)\coloneqq-\log \overline{Z}\left(\vec{\mu}\right)$; the Rényi-2 entropy is then given by the free energy cost of flipping down the boundary pinning fields in region $A$:
\begin{equation}
\overline{S_2(\rho_{A})}\simeq F_1 - F_0
\end{equation} 
where $F_1=-\log \overline{Z_1}$ and $F_0=-\log \overline{Z_0}$.
\section{Results}

\subsection{Homogeneous case}

In the homogeneous case, where all spins take the same value $j$, we can define $\beta\coloneqq\log d_j$ and write
\begin{equation}\begin{split}\label{h}
&\overline{Z}\left(\vec{\mu}\right)=\sum_{\vec{\sigma}}e^{-\beta H[\vec{\mu}](\vec{\sigma})}
\end{split}
\end{equation}
where 
\begin{equation}\begin{split}\label{Hhomo}
H[\vec{\mu}](\vec{\sigma})&=\beta^{-1}\mathcal{A}[\vec{\mu}](\vec{\sigma})= -\frac{1}{2}\left[\sum_{e_{vw}^i\in L}(\sigma_v\sigma_w -1) 
+\sum_{e_{v}^i\in \partial \gamma}(\sigma_v\mu_{e_v^i}-1)\right]+\beta^{-1}S_2({\rho_\zeta}_\downarrow)+ \mathrm{const}
\end{split}
\end{equation}
is the Hamiltonian function of a classical Ising model defined on $\gamma$,  with an additional term deriving from the bulk entropy. The quantity $\beta=\log d_j$ plays the role of an inverse temperature; the large-spins regime thus corresponds to the low temperature regime, in which the partition function is dominated by the lowest energy configuration, and the free energy becomes 
\begin{equation}
F\left(\vec{\mu}\right)=-\log \overline{Z}\left(\vec{\mu}\right)\approx \beta\min_{\vec{\sigma}}H[\vec{\mu}](\vec{\sigma})
\end{equation}
 As we are interested in a difference of free energies, we can set the constant in \eqref{Hhomo} equal to zero. With such a choice, $F_0=0$ (the minimum energy is reached when all Ising spins point up, so that the interaction terms vanish) and the Rényi-2 entropy is thus simply given by $F_1$. In the following we therefore focus on $H_1$ and, to simplify the notation, we omit the parametric dependence on $\vec{\mu}$. The analysis can be split into the two regimes corresponding to having the Ising-Hamiltonian term (deriving from link entaglement) dominant with respect to the bulk Rényi entropy (intertwiner entanglement) and vice versa.

\subsubsection{Non-dominant bulk entropy: Ryu–Takayanagi formula for homogeneous spin networks}

	In the case $S_2({\rho_\zeta}_\downarrow)=0$ the Hamiltonian is given by 
	\begin{equation}\begin{split}
	H_1(\vec{\sigma})= -\frac{1}{2}\left[\sum_{e_{vw}^i\in L}(\sigma_v\sigma_w -1)
	+\sum_{e_{v}^i\in \partial \gamma}(\sigma_v\mu_{e_v^i}-1)\right]
	\end{split}
	\end{equation}
Every pair of linked vertices with anti-parallel spins ($\sigma_v\sigma_w=-1$) carries a contribution to the energy equal to 1, and the same happens for pairs of boundary Ising-spin and pinning field ($\sigma_v\mu_v^i=-1$). The value of $H_1(\vec{\sigma})$ thus coincides with the size of the domain wall $\Sigma(\vec{\sigma})$ between the spin-up and the spin-down regions\footnote{Here the pinning fields $\vec{\mu}$ are treated at the same level of the Ising spins.}, quantified by the number of links crossing it: $|\Sigma(\vec{\sigma})|=H_1(\vec{\sigma})$. We thereby obtain
	\begin{equation}\label{RThomo}
	\overline{S_2(\rho_{A})} \simeq \log d_j \min_{\vec{\sigma}} |\Sigma(\vec{\sigma})|
	\end{equation}
	where the left hand side, in which $\log d_j$ multiplies the number of links across the minimal $\Sigma(\vec{\sigma})$, provides the \textit{area} of the latter ($\log d_j$ is in fact proportional to the area of the surface dual to a link). 
Recall that in the $\overline{Z_0}$ configuration all Ising spins point up; when switching to the $\overline{Z_1}$ one, the Ising spins in the immediate proximity of $A$ are induced to flip down. From there the spin-down region spreads, stopping when $H_1(\vec{\sigma})$, namely the size of the domain wall, is minimized. This means that \eq{RThomo} is a version of the Ryu–Takayanagi formula for homogeneous spin networks.
	
	If $S_2({\rho_\zeta}_\downarrow)$ is not null, but still negligible respect to the contribution to $H_1$ deriving from the interactions of Ising spins to each other and to pinning fields, the average Rényi-2 entropy continues to satisfy the Ryu–Takayanagi formula, with $S_2({\rho_\zeta}_\downarrow)$ a small correction to the area-law term:
	\begin{equation}\label{alaw}
	\overline{S_2(\rho_A)}\simeq\log d_j \left(\min_{\vec{\sigma}} |\Sigma(\vec{\sigma})| \right)+ S_2({\rho_\zeta}_\downarrow)
	\end{equation}
	where the minimization over $\vec{\sigma}$ is performed independently from the bulk term, and the spin-down region entering the latter is the one selected by this minimization procedure.
\subsubsection{Large bulk entropy and emergence of horizon-like regions in homogeneous spin networks}

	We now consider the case in which the bulk entanglement contribution to $H_1(\vec{\sigma})$ is comparable to that of internal and boundary edges. This happens, for example, when the bulk is in a random pure state, namely has Rényi-2 entropy given by	
		\begin{equation}
	S_2({\rho_\zeta}_\downarrow)=\log \frac{ D_j^{N}+1}{ D_j^{|\sigma_\downarrow|}+D_j^{|\sigma_\uparrow|}}
	\end{equation}
	where $\sigma_\downarrow$ ($\sigma_\uparrow$) is the region with Ising spins pointing up (down).
	For vertices of valence 4 the intertwiner dimension is $D_j=d_j=e^\beta$, and for $\beta \gg 1$ it holds that\footnote{In fact 
		\begin{equation}\begin{split}
		\frac{ e^{\beta N}+1}{e^{\beta |\sigma_\uparrow|}+ e^{\beta |\sigma_\downarrow|}} \simeq  \frac{ e^{\beta N}}{e^{\beta |\sigma_\uparrow|}(1+ e^{\beta (|\sigma_\downarrow| - |\sigma_\uparrow|)})} \simeq   e^{\beta (N-\max \{|\sigma_\uparrow|, |\sigma_\downarrow|\})}= e^{\beta \min \{|\sigma_\uparrow|,|\sigma_\downarrow|\}}.
		\end{split}
		\end{equation}} $S_2({\rho_\zeta}_\downarrow)\simeq \beta \min \{|\sigma_\uparrow|,|\sigma_\downarrow|\}$. The Hamiltonian thus takes the following form:
	\begin{equation}\begin{split}\label{random}
	H_1(\vec{\sigma})=-\frac{1}{2}\left[\sum_{e_{vw}^i\in L}(\sigma_v\sigma_w -1)
	+\sum_{e_{v}^i\in \partial \gamma}(\sigma_v\mu_{e_v^i}-1)\right] + \min \{|\sigma_\uparrow|,|\sigma_\downarrow|\}
	\end{split}
	\end{equation}
	In this case, the minimization of $H_1(\vec{\sigma})$, namely the behaviour of the minimum-size domain wall, strongly depends on the bulk entanglement contribution. We illustrate such an effect with an example; in particular, we show that the bulk entanglement can remove the degeneration between equal-energy configurations (i.e. equal-size domain walls) and, when sufficiently high in a given region, forces the domain wall to stay outside of it.
	
\begin{figure}[t]
		\begin{minipage}[t]{0.48\linewidth}
			\centering
			\includegraphics[width=0.8\linewidth]{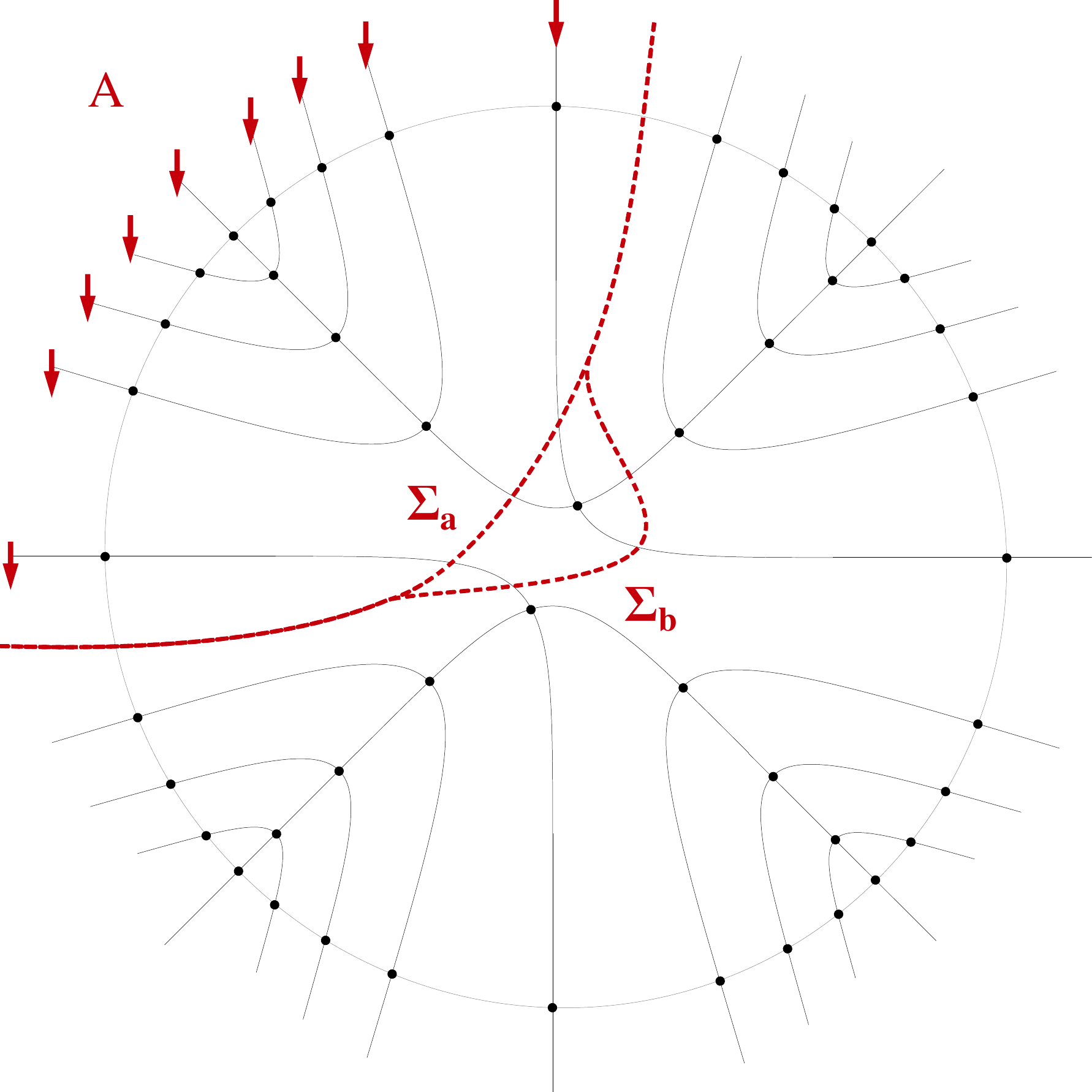}\caption{The minimum of $H_1$ is degenerate; the corresponding surfaces $\Sigma_a$ and $\Sigma_b$, with area $|\Sigma_a|=|\Sigma_b|=5$, are shown by dashed red lines.}
			\label{deg}
		\end{minipage}
		\hspace{0.5cm}
		\begin{minipage}[t]{0.48\linewidth}
			\centering
			\includegraphics[width=0.8\linewidth]{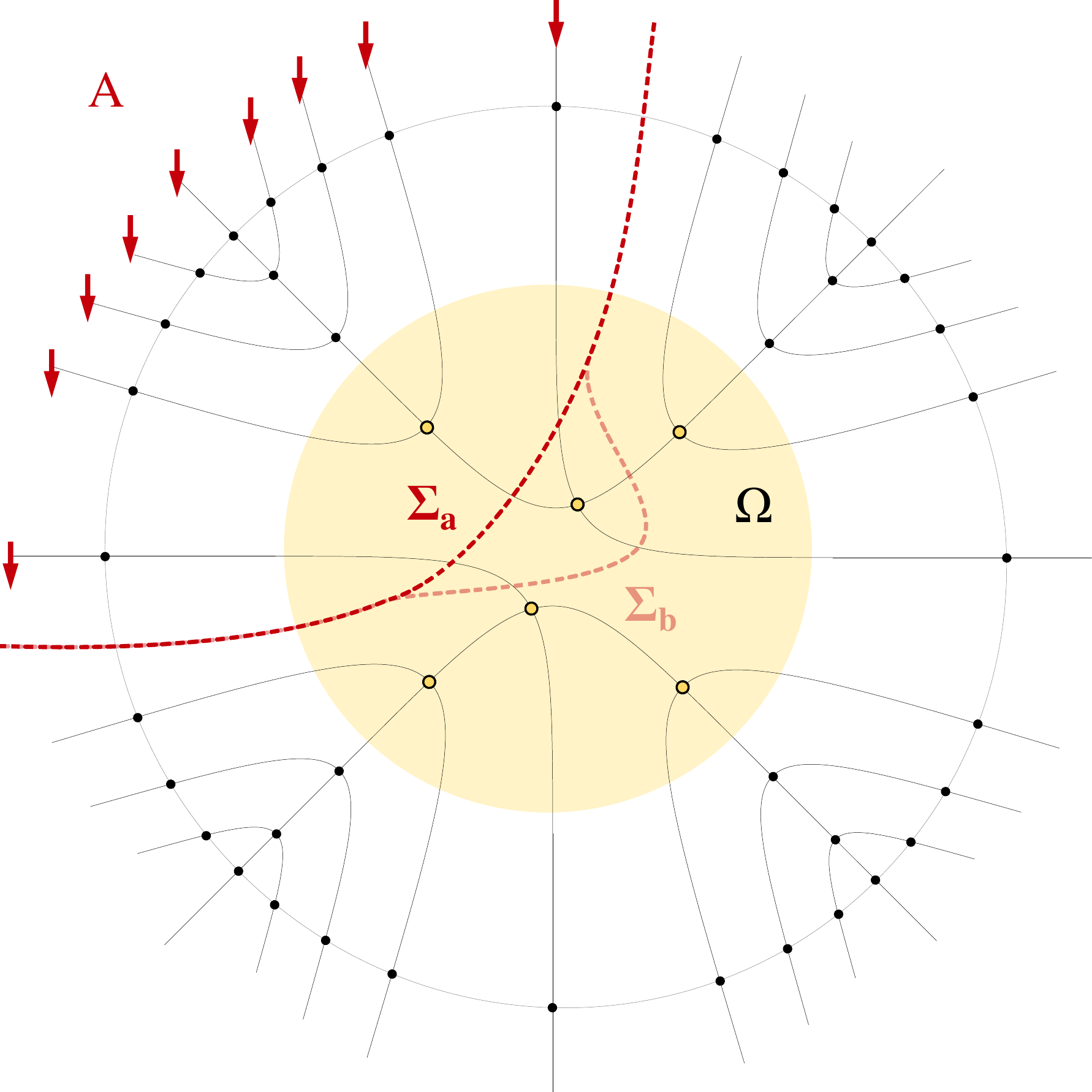}\caption{If intertwiner entanglement is present in a region $\Omega$ of the bulk (highlighted in yellow in the figure, as well as the vertices within it), the degeneracy of the minimal energy is removed. In fact, $|\Sigma_a|=6$ and $|\Sigma_b|=7$.}
			\label{deg_removed}
		\end{minipage}
	\end{figure}

	Consider the homogeneous graph depicted in figure~\ref{deg}. Since it is made of four-valent vertices, the dimension of the intertwiner degrees of freedom is equivalent to that of the edges: $D_j=d_j=2j+1$. The boundary region $A$ we look at, together with its pointing-down pinning fields, is illustrated in figure \ref{deg}. 
	\begin{figure}[t]\includegraphics[width=0.384\linewidth]{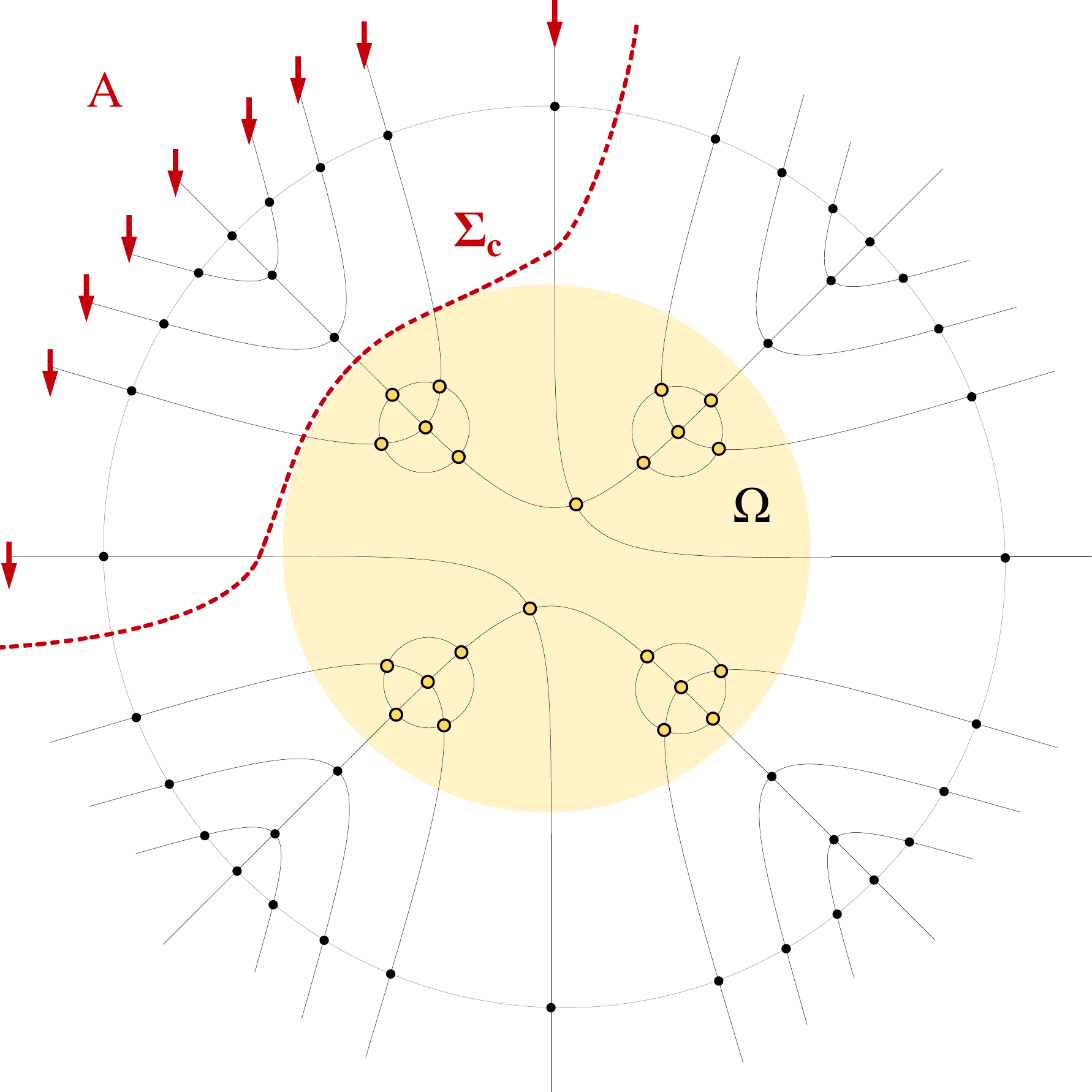}\caption{By increasing the dimension of the bulk disk $\Omega$ via refinement of vertices, the minimal-energy surface $\Sigma_c$ is prevented from entering it.}\label{out_bulk}
	\end{figure}	
	We first assume that the bulk state is a separable state of all intertwines, therefore its entanglement entropy contribution is zero and $H(\vec{\sigma})=|\Sigma(\vec{\sigma})|$. The minimal energy $H_1=4$ is then reached by two configurations (denoted as $\vec{\sigma}_a$ and $\vec{\sigma}_b$) whose corresponding surfaces $\Sigma_a \coloneqq \Sigma(\vec{\sigma}_a)$ and $\Sigma_b \coloneqq \Sigma(\vec{\sigma}_b)$ are depicted in figure \ref{deg}.
	
	We now switch on the intertwiner correlations in a bulk disk $\Omega$, which is illustrated in figure \ref{deg_removed}; in particular, we assume $\Omega$ to be in a random pure state $\ket{\zeta_{\Omega}}$, with the complementary part of the bulk being in a direct product state:
		\begin{equation}
		\ket{\zeta}=\ket{\zeta_{\Omega}}\otimes \bigotimes_{v:~r_v>1}\ket{\xi_v}
	\end{equation}
The Hamiltonian thus takes the form (see \eq{random})
	\begin{equation}\begin{split}\label{randomOmega}
	H_1(\vec{\sigma})=-\frac{1}{2}\left[\sum_{e_{vw}^i\in L}(\sigma_v\sigma_w -1)
	+\sum_{e_{v}^i\in \partial \gamma}(\sigma_v\mu_{e_v^i}-1)\right] + \min \{|\Omega_\uparrow|,|\Omega_\downarrow|\}
	\end{split}
	\end{equation}
where $\Omega_\downarrow\coloneqq  \Omega \cap \sigma_\downarrow$ and $\Omega_\uparrow \coloneqq  \Omega \cap \sigma_\uparrow$. For the two configurations $\vec{\sigma_a}$ and $\vec{\sigma}_b$ we then have that
\begin{align}
&H_1(\vec{\sigma_a})= 4 + \beta^{-1}\log \frac{e^{9\beta}+1}{e^{2\beta}+e^{7\beta}}\approx 4 + \beta^{-1}\log e^{2\beta}=6\\&H_1(\vec{\sigma_b})= 4 + \beta^{-1}\log \frac{e^{9\beta}+1}{e^{3\beta}+e^{6\beta}}\approx 4 + \beta^{-1}\log e^{3\beta}=7 \quad .
\end{align}
When the bulk correlations are switched on, the degeneracy of the minimum of $H_1$ is thus removed. Note that the lowest-energy configuration,  $\vec{\sigma_a}$, has a domain wall that  enters the bulk disk $\Omega$. However, upon increasing the dimension of the latter, the domain wall ends up being pushed out of it. To show this, we just have to refine vertices of the bulk disk $\Omega$ as shown in figure \ref{out_bulk}. The Ising energy of $\vec{\sigma_a}$ then increases to $H_1(\vec{\sigma_a})=14$, and the domain wall settles down outside $\Omega$, with  $H_{1\mathrm{min}}=8$.

\subsection{Inhomogeneous case}
We consider here the more general case in which each link of the graph carries a different spin\footnote{We still have a fixed spin assignment, though, and no superposition of fixed-spin states, whose analysis is left for future work.}. From the Ising model point of view, this means that we are considering Ising spins with inhomogeneous couplings. Nevertheless, in order to define a Boltzmann weight for the partition function, we introduce a uniform $\beta$ to be understood as an average inverse temperature, i.e.  $\beta=\log d$ where $d$ is the average edge dimension; then $\overline{Z}(\vec{\mu})=\sum_{\vec{\sigma}}e^{-\beta H[\vec{\mu}](\vec{\sigma})}$ with
\begin{equation}\begin{split}
&H[\vec{\mu}](\vec{\sigma})= -\frac{1}{2}\sum_{e_{vw}^i\in L}(\sigma_v\sigma_w -1) J_{vw}^i
-\frac{1}{2}\sum_{e_{v}^i\in \partial \gamma}(\sigma_v\mu_{e_v^i}-1) J_{v}^i + \beta^{-1}S_2({\rho_\zeta}_\downarrow)
\end{split}
\end{equation}
where $J_{vw}^i\coloneqq\frac{\log d_{j_{vw}^i}}{\beta}$ and $J_{v}^i\coloneqq \frac{\log d_{j_v^i}}{\beta}$ are the (normalised) strengths of the interaction. The large-spins regime thus corresponds to the low temperature regime, in which the partition function is dominated by the minimum energy configuration, and $F=-\log \overline{Z}\simeq \beta H_{\text{min}}$. Note also that, similarly to what happens for the homogeneous case, $F_0=0$; in fact, when $\mu_{e}=+1~ \forall e \in \partial \gamma$, the lowest energy configuration is that with all Ising spins pointing up, for which all terms in $H_0$ are zero. When, instead, $\mu_{e}=-1$ for $e\in A$ and $\mu_{e}=+1$ for $e\notin A$ (boundary condition for $F_1$), a spin-up region (with external boundary $\overline{A}$) and a spin-down region (with external boundary $A$) arise, and the domain wall settles down in order to minimize $H_1$. 
\subsubsection{Non-dominant bulk entropy: Ryu–Takayanagi formula for inhomogeneous spin networks}

For null bulk entanglement entropy we have the Ising Hamiltonian
	\begin{equation}\begin{split}\label{H}
	H_1(\vec{\sigma})&= -\frac{1}{2}\sum_{e_{vw}^i\in L}(\sigma_v\sigma_w -1) J_{vw}^i
	-\frac{1}{2}\sum_{e_{v}^i\in \partial \gamma}(\sigma_v\mu_{e_v^i}-1) J_{v}^i
	\end{split}
	\end{equation}
	that provides the \textit{area} of the domain wall, determined by both combinatorial and dimensional properties of the entanglement graph. In fact, it not just the number of open-edges/links that matters: every open-edge $e_v^i$ (link $e_{vw}^{i}$) is weighted by a factor $J_{v}^i$ ($ J_{vw}^i $) proportional to (the logarithm of) its dimension (which, in turns, gives the area of the surface topologically dual to the link). An analogue of the Ryu-Takayanagi formula therefore holds and, due to the (quantum) discrete geometric nature of the degrees of freedom carried by the entanglement graphs, it involves a properly geometric notion of area (i.e. the spin degrees of freedom concur to the definition of the discrete geometry, and the discrete metric is not simply given by the graph distance). Similarly to the homogeneous counterpart, the presence of small bulk entanglement entropy represents a (not negligible, but small) correction to the area term:
	\begin{equation}
	\overline{S_2(\rho_A)}\simeq\log d_j \left(\min_{\vec{\sigma}} |\Sigma(\vec{\sigma})| \right)+ S_2({\rho_\zeta}_\downarrow)
	\end{equation}
	with $|\Sigma(\vec{\sigma})|$ given by \eq{H}.
	\subsubsection{Larger bulk entropy and emergence of horizon-like regions in inhomogeneous spin networks}
	When the contribution of the bulk entanglement entropy is not small with respect to the Ising part, we need to minimize the whole Hamiltonian:
	\begin{equation}\begin{split}\label{H1}
	H_1(\vec{\sigma})&=  -\frac{1}{2}\sum_{e_{vw}^i\in L}(\sigma_v\sigma_w -1) J_{vw}^i
	-\frac{1}{2}\sum_{e_{v}^i\in \partial \gamma}(\sigma_v\mu_{e_v^i}-1) J_{v}^i +\beta^{-1}S_2({\rho_\zeta}_\downarrow) \quad .
	\end{split}
	\end{equation}
	Note that an internal link $e_{vw}^i$ carries a contribution ($J_{vw}^i$) to the energy only if the Ising spins $\sigma_v$ and $\sigma_w$ are misaligned; a boundary edge in $A$ carries a contribution ($J_{v}^i$) only if the Ising spin $\sigma_v$ points up, while a boundary edge in $\overline{A}$ carries a contribution ($J_{v}^i$) only if the Ising spin $\sigma_v$ points down. As a result, the first two terms of $H_1$ are minimized by the configuration whose spin-down region $\sigma_\downarrow$ has external boundary $A$ and internal boundary $|\Sigma(\vec{\sigma})|$ of smallest possible size. Then, the region $\sigma_\downarrow$ contributes to $H_1$ with the ($\beta$-rescaled) Rényi-2 entropy of its reduced bulk state  ${\rho_\zeta}_\downarrow$. In the end, the minimization of $H_1$ is achieved when the spin-down region $\sigma_\downarrow$ has the minimum possible area and volume correlations. 
	
\begin{figure}[t]\includegraphics[width=0.3\linewidth]{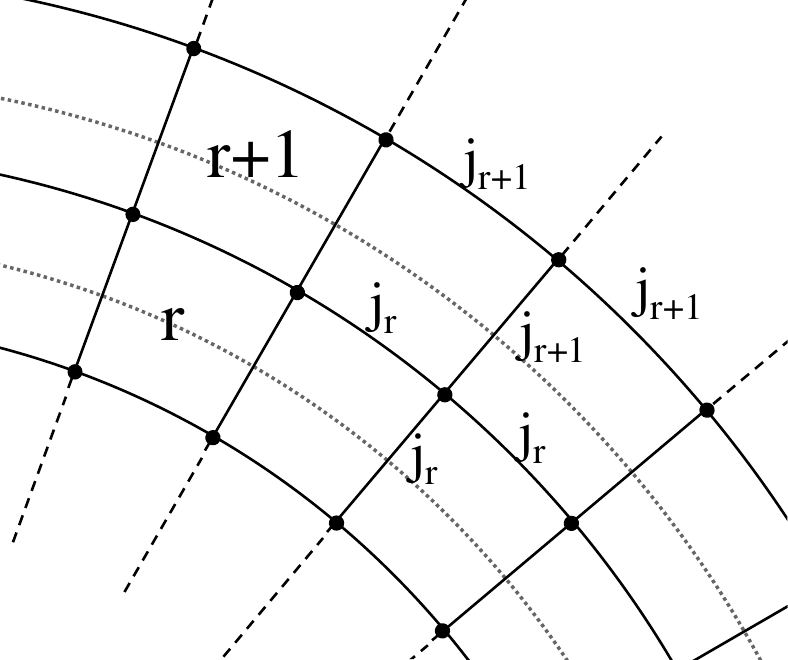}\caption{Each shell $r$ crosses radial links carrying spin $j_r$; vertices between shells $r$ and $r+1$ recouple three spins $j_r$ (one on the radial inward direction, two on the edges tangent to shell $r$) and one spin $j_{r+1}$ (outward radial direction). }\label{spherical}
	\end{figure}

	We illustrate the properties of this mechanism with an example. Consider the (spherically symmetric) graph of figure\ti\ref{spherical}, with a radial gradient of edge spins: $j_{r+1}>j_r$. 
	We assume that the bulk is in a random pure state inside a disk $\Omega$ of radius $R$ and in a product state outside. Therefore,
	\begin{equation}\label{S}
	S_2({\rho_\zeta}_\downarrow)=\log \frac{\prod_{v\in \Omega} D_{\text{j}_v}+1}{\prod_{v\in   \Omega_\downarrow } D_{\text{j}_v}+\prod_{v\in \Omega_\uparrow} D_{\text{j}_v}}
	\end{equation}
	We express the intertwiner dimensions $D_{\text{j}_v}$ in \eq{S} as local intertwiner inverse temperatures $\beta'_v=\log D_{\text{j}_v}$; we then consider the large spins regime and assume that the variance of intertwiner dimensions within $\Omega$ is small, i.e. $\beta'_v \approx \beta'$ for all $v\in \Omega$. Then \eq{S} simplifies to\footnote{In the large spins regime
		\begin{equation}
		\log \left( \frac{\prod_{v\in \Omega} D_{\text{j}_v}+1}{\prod_{v\in   \Omega_\downarrow } D_{\text{j}_v}+\prod_{v\in \Omega_\uparrow} D_{\text{j}_v}}\right)\approx \log \left( \frac{\prod_{v\in \Omega}e^{\beta'_v}}{\prod_{v\in   \Omega_\downarrow } e^{\beta'_v}+\prod_{v\in \Omega_\uparrow} e^{\beta'_v}}\right)=\sum_{v\in \Omega} \beta'_v- \log \left( \prod_{v\in   \Omega_\downarrow } e^{\beta'_v}+\prod_{v\in \Omega_\uparrow} e^{\beta'_v}\right)
		\end{equation}
		If we assume that $\beta'_v \approx \beta'$ for all $v$ then
		\begin{equation}
		\prod_{v\in   \Omega_\downarrow } e^{\beta'_v}+\prod_{v\in \Omega_\uparrow} e^{\beta'_v} \approx    e^{\beta'\max \{|\Omega_\uparrow|,|\Omega_\downarrow|\}}\left[1 + e^{-\beta'\left(\max \{|\Omega_\uparrow|,|\Omega_\downarrow|\}- \min \{|\Omega_\uparrow|,|\Omega_\downarrow|\}\right)}\right] \approx    e^{\beta'\max \{|\Omega_\uparrow|,|\Omega_\downarrow|\}}
		\end{equation}
		and we get
		\begin{equation}
		\sum_{v\in \Omega}\beta'_v - \log \left( \prod_{v\in   \Omega_\downarrow } e^{\beta'_v}+\prod_{v\in \Omega_\uparrow} e^{\beta'_v}\right) \approx \beta'\left(\Omega - \max \{|\Omega_\uparrow|,|\Omega_\downarrow|\}\right) =\beta'\min \{|\Omega_\uparrow|,|\Omega_\downarrow|\}
		\end{equation}
	} $S_2({\rho_\zeta}_\downarrow)= \beta' \min\{|\Omega_\uparrow|,|\Omega_\downarrow|\}$  and 
	\begin{equation}\label{happrox}\begin{split}
	H(\vec{\sigma})\approx -\frac{1}{2}\sum_{e_{vw}^i\in L}(\sigma_v\sigma_w -1) {J_{vw}^i}
	-\frac{1}{2}\sum_{e_{v}^i\in \partial \gamma}(\sigma_v\mu_{e_v^i}-1){J_{v}^i} + \frac{\beta'}{\beta}\min\{|\Omega_\uparrow|,|\Omega_\downarrow|\}
	\end{split}
	\end{equation}
	Note that, with respect to the homogeneous case of \eq{randomOmega}, the smallest number of aligned spins in the bulk region $\Omega$ is now weighted by the ratio of the intertwiner inverse temperature $\beta'$ to the link inverse temperature $\beta$.
	
\begin{figure}[t]
	\begin{minipage}[t]{0.48\linewidth}
		\centering
		\includegraphics[width=0.8\linewidth]{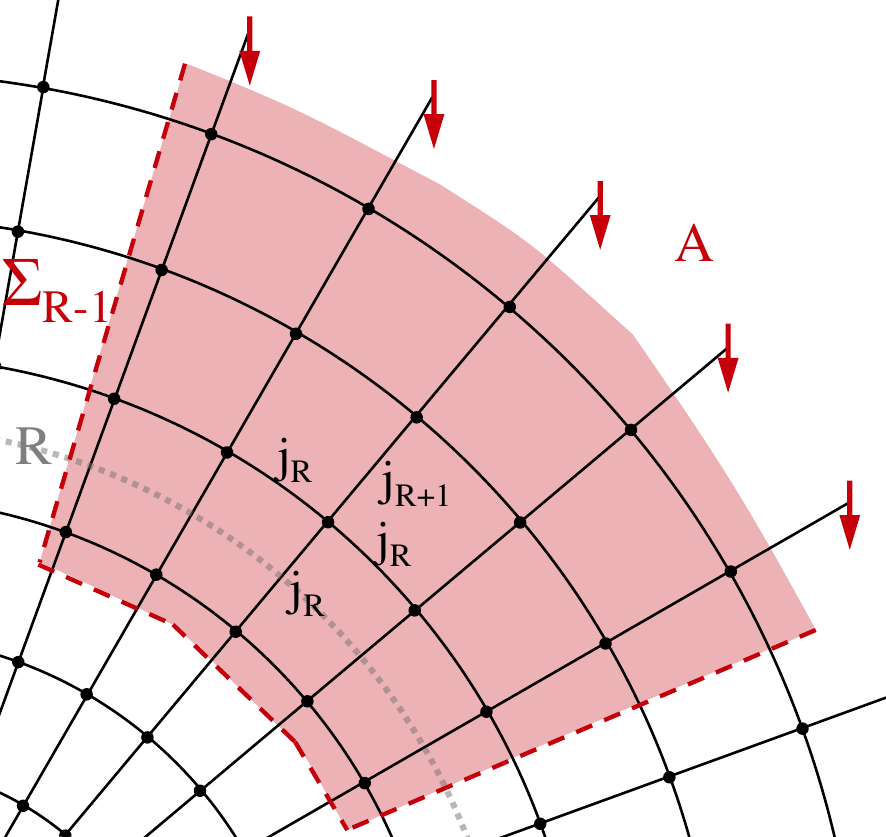}\caption{In absence of bulk entanglement, the Ising domain wall (the dashed red line) moves towards the center of the spherical geometry.}
		\label{in}
	\end{minipage}
	\hspace{0.5cm}
	\begin{minipage}[t]{0.48\linewidth}
		\centering
		\includegraphics[width=0.8\linewidth]{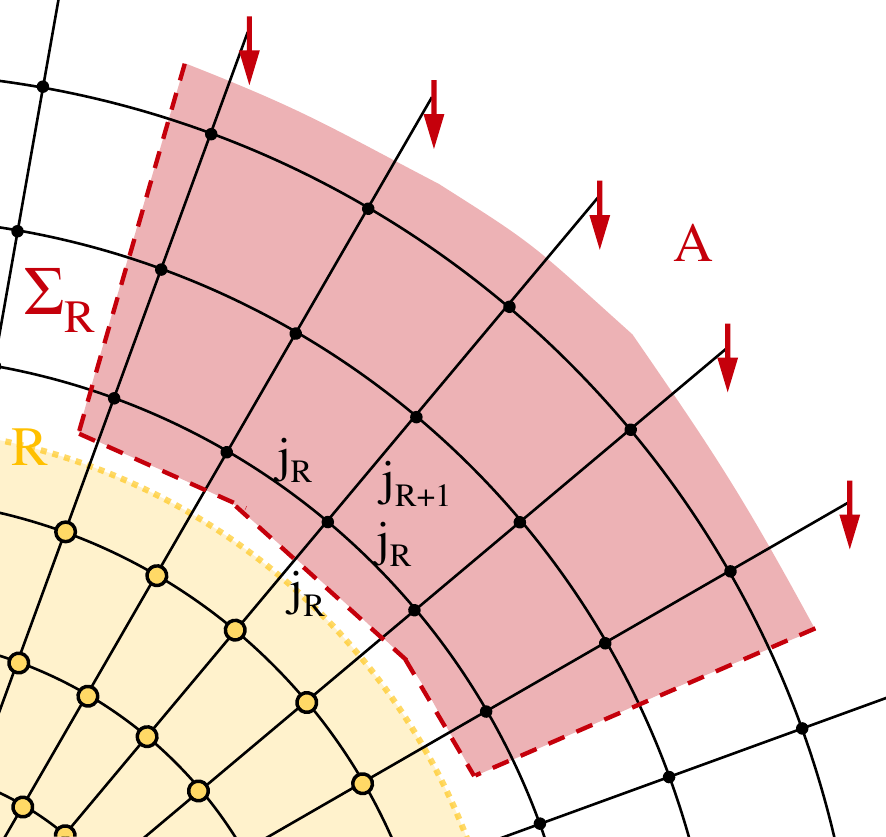}\caption{When a bulk disk of radius $R$ is in a random pure state, the Ising domain wall is prevented from entering it.}
		\label{out}
	\end{minipage}
\end{figure}

	We are going to show how the presence of intertwiner entanglement in a disk $\Omega$ of the spherically symmetric graph in figure\ti\ref{in} affects the entanglement entropy of a portion $A$ of the boundary. Note that, as we want $D_{j_r j_r j_r j_{r+1}}=3j_r-j_{r+1}+1>1$ inside $\Omega$, we must have $j_r<j_{r+1}<3j_r$ for $r\le R$.
Let $ \mathcal{A}(r)$ be the Ising action of a configuration whose domain wall $\Sigma_r$ lies between shell $r$ and shell $r-1$ (see figure~\ref{in}). 
When the bulk entanglement is not present, we have that
\begin{equation}\label{Ar}
 \mathcal{A}(r)=\left(|A|+2\right) \log d_{j_{r}} + 2 \sum_{k=r+1}^{r_{\mathrm{max}}}\log d_{j_{k}}
\end{equation}
The minimal-energy surface drops from shell $r+1$ to shell $r$ if $\mathcal{A}(r+1)> \mathcal{A}(r)$. By using \eqref{Ar}, the latter becomes
\begin{equation}
 |A| \log d_{j_{r+1}}> (|A|+2) \log d_{j_r} \quad ,
\end{equation}
which is satisfied by 
\begin{equation}
d_{j_{r+1}}>   d_{j_{r}}^{\frac{|A|+2}{|A|}} \quad ,
\end{equation}
that, for $|A|\gg 1$, is always true. We therefore have that, in absence of bulk entanglement, the minimal-energy surface moves toward the innermost shells.

When switching on the bulk entanglement within $\Omega$ (specifically, when assuming that $\Omega$ is in a random pure state), the value of the Ising action for the domain wall at $r=R$ is no more given by \eq{Ar}. Instead we have
\begin{equation}
 \mathcal{A}(R)=\left(|A|+2\right) \log d_{j_{R}} + 2 \sum_{k=R+1}^{r_{\mathrm{max}}}\log d_{j_{k}} + |A|\log \left(\frac{d_{j_{R}}+d_{j_{R+1}}}{2}\right)
\end{equation}
where we used the fact that $D_{j_r j_r j_r j_{r+1}}=3j_r-j_{r+1}+1=\frac{d_{j_{r}}+d_{j_{r+1}}}{2}$. The condition for the domain wall to enter $\Omega$, i.e.  $\mathcal{A}(R+1)-\mathcal{A}(R)>0$, thus leads to
\begin{equation}
  |A| \log d_{j_{R+1}}> (|A|+2) \log d_{j_{R}} +|A|\log \left(\frac{d_{j_{R}}+d_{j_{R+1}}}{2}\right)
\end{equation}
which can be written as follows:
\begin{equation}\label{never}
d_{j_{R+1}}\left(2- d_{j_R}^{\frac{|A|+2}{A}}\right)>d_{j_R}^{1+\frac{|A|+2}{A} } \quad .
\end{equation}
Since $d_j\ge 2$ for $j>0$, the left hand side of \eqref{never} is negative, and \eqref{never} is therefore never satisfied: the minimal-energy surface is prevented from entering the disk $\Omega$, as shown in figure\ti\ref{out}. We thus found that the presence of (large) intertwiner entanglement within the disk $\Omega$ makes its boundary (the shell of radius $R$) a horizon-like region.

\section{Conclusions and outlook}

We have studied, for spin network states corresponding to random tensor networks, how the Rényi-2 entropy of the boundary is affected by the bulk data, specifically by its combinatorial structure and by the quantum correlations among the intertwiners. We relied on random tensor network techniques (specifically, we adapted to our framework the ones of \cite{Hayden:2016cfa}). This led to the following scenario: in absence of intertwiner correlations, randomizing over vertex wave-functions with uniform probability measure maps the entropy calculation to the free energy of an Ising model living on the spin network graph, in which the strength of interaction depends only on the dimension of the graph degrees of freedom, and the boundary entropy is given by the free energy cost of shifting the Ising domain wall. 

We then found that, when the intertwiner entanglement is not present, the boundary entropy of our spin network states follows the Ryu–Takayanagi formula, with the Ryu–Takayanagi surface carrying a clear quantum-geometry interpretation. The presence of quantum correlations among the intertwiners adds a term to the Ising Hamiltonian, which is given by the entropy of the bulk state. This term affects the position of the Ising domain wall and thus the entropy of the boundary. In particular, for small values of the bulk entropy we recover the Ryu–Takayanagi formula with a bulk-induced correction. We also showed that, when a bulk region with high Rényi-2 entropy is present, the Ising domain wall cannot enter it. In other words, such bulk region behaves, from a purely information-theoretic perspective, exactly like an event horizon.

Our results can be generalised in several directions. The next steps will be to analyse in detail the behaviour of the domain wall in the case of superpositions of spin network states for given combinatorial structure, and of superpositions of the combinatorial structure of the associated graph. 

The same results are also a solid basis for many interesting developments. The role of the dimension of the spin/intertwiner degrees of freedom and the entanglement structure of the intertwiners need to be explored further too. In a more physically important direction, from the point of view of (classical an) quantum gravity, we are interested in exploring further the conditions of the emergence of bulk regions in which the domain wall cannot access. This could pave the way to a more comprehensive information theoretic characterization of black hole-like regions in fundamental quantum gravity (see \cite{Livine:2005mw,Anza:2016fix,Anza:2017dkd} for other potentially useful results in this direction).
\acknowledgments
The authors would like to thank the quantum gravity group at LMU for useful discussions and comments. E. Colafranceschi also thanks the Ludwig Maximilian University of Munich for the hospitality.
D. Oriti acknowledges funding from the Deutsche Forschung Gemeinschaft (DFG), via the research grant OR432/3-1. G. Chirco acknowledges funding from INFN, via the INFN Fellowship in Theoretical Physics, hosted by the Physics Department ``Ettore Pancini'' at the University of Naples Federico II. 
\bibliographystyle{jhep}
\bibliography{biblio}

\end{document}